\documentclass[traditabstract]{aa}  
\usepackage{graphicx}
\usepackage{multirow}
\usepackage{verbatim}
\usepackage{lscape,textcomp} 
\usepackage{txfonts}
\usepackage{natbib}
\usepackage{acronym,amsfonts} 

\bibpunct{(}{)}{;}{a}{}{,}\newcommand{\hh}{H$_{2}$}
\newcommand{\kms}{km~s$^{-1}$}
\newcommand{\nhh}{$N(\mathrm{H}_{2})$}
\newcommand{\tten}[1]{\times 10^{{#1}}}
\newcommand{\msun}{$M_{\odot}$}

\acrodef{lmc}[LMC]{Large Magellanic Cloud}
\acrodef{smc}[SMC]{Small Magellanic Cloud}

\begin{document}
   \title{Millimeter dust emission compared with other mass estimates in N11 molecular clouds in the LMC}
   \author{C. N. Herrera\inst{1,}\inst{2},
   M. Rubio\inst{2},
   A. D. Bolatto\inst{3},
   F. Boulanger\inst{2},        
        F. P. Israel\inst{4},
         \and 
         F. T. Rantakyr\"o\inst{5}     
          }
   \institute{   
   Institut d\textquotesingle Astrophysique Spatiale (IAS), UMR8617, Universit\'e Paris-Sud 11,
	      B\^{a}timent 121, 91405 Orsay Cedex, France.
        \and
	      Departamento de Astronom\'ia, Universidad de Chile,
              Casilla 36-D, Las Condes, Santiago Chile.
         \and
              Department of Astronomy and Laboratory for Millimeter-Wave Astronomy, University of Maryland, College Park, MD 20742.
          \and
              Sterrewacht Leiden, Leiden University, PO Box 9513, 2300 RA, Leiden, The Netherlands. 
         \and
              Gemini Observatory, Southern Operations Center, c/o AURA, Casilla 603, La Serena, Chile.              
              }


\abstract
{
CO and dust emission at millimeter wavelengths are independent tracers of cold interstellar matter, which have seldom been
compared on the scale of  giant molecular clouds (GMCs) in other galaxies. In this study, and for the first time in the Large Magellanic Cloud (LMC), we compute the molecular cloud masses from the millimeter emission of the dust and compare them with the masses derived from their CO luminosity and virial theorem. We present  CO (J$=$1--0 and J$=$2--1)  and 1.2 mm continuum observations of the N11 star forming region in the LMC obtained with the SEST telescope and the SIMBA bolometer, respectively. We use the CO data to identify individual molecular clouds and measure their physical properties (CO luminosity, size, line width and virial masses). The correlations between the properties of the N11 clouds are in agreement with those found in earlier studies in the LMC that sample a larger set of clouds and a larger range of cloud masses. For the N11 molecular clouds, we compare the masses estimated from the CO luminosity ($X_{\rm CO}\,L_{\rm CO}$), the virial theorem ($M_{\rm vir}$) and the millimeter dust luminosity ($\mathcal{L}_{\rm 1.2mm}({\rm dust})$). The measured ratios $L_{\rm CO}/M_{\rm vir}$ and $\mathcal{L}_{\rm 1.2mm}({\rm dust})/M_{\rm vir}$ constrain the $X_{\rm CO}$ and $\mathcal{K}_{\rm dust}$ (dust emissivity at 1.2 mm per unit gas mass) parameters as a function of the virial parameter $\alpha_{\rm vir}$. The comparison between the different mass estimates yields a $X_{\rm CO}$-factor of 8.8$\pm$3.5~$\tten{20}$ cm$^{-2}$~(K~km~s$^{-1}$)$^{-1}\,\alpha_{\rm vir}^{-1}$ and a $\mathcal{K}_{\rm dust}$  parameter of 1.5$\pm$0.5~$\times 10^{-3}~{\rm cm^2\,g^{-1}}\,\alpha_{\rm vir}$. We compare our N11 results with a similar analysis for molecular clouds in the Gould's Belt in the solar neighborhood. We do not find in N11 a large discrepancy between the dust millimeter and virial masses as reported in earlier studies of molecular clouds in the Small Magellanic Cloud. The ratio between $\mathcal{L}_{\rm 1.2mm}$ and $M_{\rm vir}$ in N11 is half of that measured for Gould's Belt clouds, which can be accounted for by a factor of two lower gas-to-dust mass ratio, as the difference in gas metallicities. If the two samples have similar $\alpha_{\rm vir}$ values, this result implies that their dust far-IR properties are also similar. 

}

\authorrunning{C. N. Herrera, et al.}

\maketitle

%


\section{Introduction}

Giant Molecular Clouds (GMCs) are the nursery of massive stars. They are of prime importance to the evolution of their host galaxies, 
but there are still uncertainties on how to best estimate their masses. At their low temperatures ($\sim10-50$~K), the molecular masses cannot be measured through the emission of the H$_2$ molecule because its first rotational transition arises from an energy level at a temperature of 510~K. 
Generally in galaxies, low-J transitions from the second most abundant molecule, CO, are the tracer of the molecular mass. 
Dust emission is an alternative tracer of the gas mass, which has been used to estimate dust masses \citep{draine07b,galametz11} and to calibrate the $X_{\rm CO}$ conversion factor from CO luminosity to H$_2$ mass  \citep{israel97,leroy11}. In the Milky Way and nearby galaxies, where GMCs are spatially resolved, the $X_{\rm CO}$-factor has been also estimated using the relationships between cloud properties assuming virial equilibrium \citep{solomon87,bolatto08}. 

CO emission does not have a one-to-one correspondence with H$_2$. Observations in the Galaxy \citep{grenier05,abdo10, planck11a19} and models of photodissociation regions (PDRs) \citep[e.g.][]{wolfire10} show that CO does not trace the molecular gas over the full extent of clouds. CO-dark H$_2$ gas was also observed in the Large Magellanic Cloud (LMC) with dust observations from Spitzer and Herschel \citep{bernard08, romanduval10,galliano11}.  In the Small Magellanic Cloud (SMC), \citet{rubio04} and later \citet{bot07,bot10} reported differences up to a factor $\sim$10 between GMCs masses derived from the dust emission (based on 870~$\mu$m and 1.2~mm emission) and the virial theorem. Additional evidence for H$_2$ gas without CO emission is provided by observations of the [\ion{C}{ii}]$\lambda$158$\mu$m line emission in Magellanic irregular galaxies \citep{madden97, israel11} and the Milky Way \citep{velusamy10}.
PDR models show that the column density threshold for CO to become the main carbon species has a much stronger dependence on the metallicity than that for the gas to become molecular \citep[e.g.][]{wolfire10}. This is because CO is shielded from photodissociation by dust while H$_2$ is self-shielded. Whereas in the Galaxy there is a contribution to CO emission from diffuse molecular gas \citep{liszt10}, in low metallicity galaxies the CO may predominately come from dense, high column density clumps embedded in a diffuse molecular envelop with weak or no CO emission \citep{israel88,lequeux94}. 
 
The motivation of this paper is to compare estimates of the molecular cloud masses in a low metallicity environment, the LMC. The LMC is at 50 kpc \citep{persson04}. It presents a metallicity, measured for \ion{H}{ii} regions, of 12+log(O/H)=8.37 \citep{keller06}, that corresponds to $Z$$\simeq$$0.5~Z_{\sun}$. Extensive CO(1--0) surveys have been done in the Magellanic Clouds. In the LMC, the first one done by \citet{cohen88} and subsequently the SEST \citep{israel93}, NANTEN \citep{fukui08} and MAGMA \citep{hughes10,wong11} surveys gave a large set of GMCs. Also, several detailed studies have been done in the LMC (e.g. \citealt{garay02} for molecular complex No. 37; \citealt{israel03} for N11; \citealt{johansson98}, \citealt{kutner97}, \citealt{pineda09} for 30Dor).  All of these studies report a CO-to-H$_2$ conversion factor estimated from the comparison between the virial and CO luminosity masses of the molecular clouds larger than in the Galaxy.
 
This paper is based on CO(J$=$1--0 and J$=$2--1) and 1.2~mm continuum observations of the N11 star-forming region in the LMC. We present the observations and compare the CO luminosity of molecular clouds with their masses estimated with the virial theorem and the millimeter dust emission. Unlike what has been done with far-IR observations from Spitzer and what is now being done with Herschel \citep{bernard08, galliano11, leroy11}, the computation of molecular masses from the millimeter emission of the dust is not very sensitive to the dust temperature because it is on the Rayleigh-Jeans spectral side of the energy distribution. Further, the comparison between dust and virial masses has seldom been performed because the molecular clouds have to be resolved to estimate virial masses. With single dish telescopes and space IR observations (Spitzer and Herschel) this is possible only in the Milky Way and the Magellanic Clouds. This paper extends the earlier studies on the SMC done by \citet{rubio04} and \citet{bot07,bot10} to intermediate metallicities in the LMC. Soon with ALMA, this can be extended to more distant galaxies in the Local group.

The paper is organized as follows. In Section 2, we present observations of CO in the J$=$2--1 transition line, new data in the transition J$=$1--0, and a continuum map at 1.2~mm. We identify the N11 molecular clouds using the CO(2--1) data and derive their physical properties (CO luminosity, size, line-width and virial mass) in Section 3. In Section 4, we derive millimeter dust fluxes from the continuum observations. In Section 5, we present the three estimations of molecular masses, and the comparison between them. We show for which values of the $X_{\rm CO}$-factor and the millimeter dust emissivity per hydrogen, as a function of the virial parameter of the clouds, the masses are identical. Finally, Section 6 presents the conclusions.

%

\section{Observations in the N11 region}

\subsection{The N11 star-forming region}
\begin{figure}[ht]
   \centering
   \includegraphics[width=8cm]{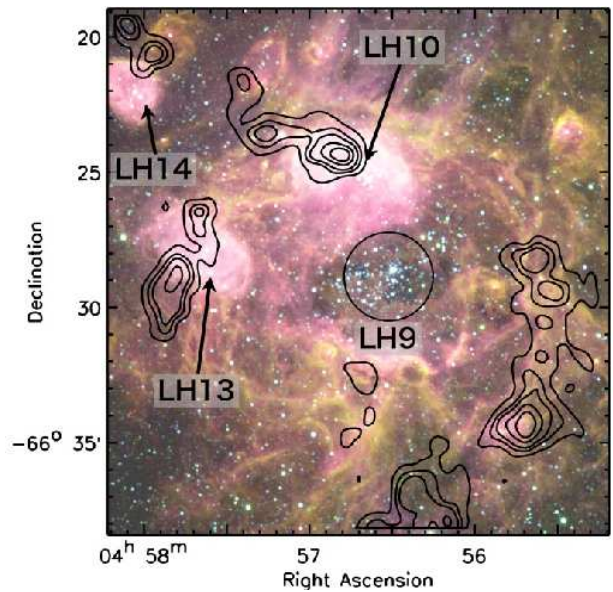}
      \caption{Image obtained from the combination of hydrogen, sulfur and oxygen bandpasses (Magellanic Clouds Emission-Line Survey MCELS, C. Aguilera, C. Smith and S. Points/NOAO/AURA/NSF).  Contours correspond to the CO(2--1) line emission. We indicate the position of the OB associations in N11, LH9 in the central cavity of N11, LH10 in the nebula N11B, LH13 in N11C and LH14 in N11E.} \label{fig:opt}
   \end{figure}

N11 \citep{henize56} is the second largest and brightest \ion{H}{ii} region in the LMC, and hosts several stellar clusters (see Fig.~\ref{fig:opt}). Its main feature is a cavity with an inner diameter of $\sim$700\arcsec\ ($\sim$170 pc). In its center there is the OB association LH9 \citep{lucke70} while in its northeast part there is the N11B nebula which is associated with another stellar cluster, LH10. The clusters are 7 and 3~Myr old, respectively \citep{mokiem07}.  \citet{parker92} show that most of the ionizing flux in N11 comes from these two clusters. IR observations \citep{barba03} reveal that the star formation is on-going in N11B. \citet{israel03} presented CO(1--0) observations of N11 with a resolution of FWHM$=$45\arcsec. They identified molecular clouds and computed their physical properties. Recently, \citet{israel11} showed that even though the [\ion{C}{ii}]$\lambda$158$\mu$m and CO emission lines present similar spatial distributions, the [\ion{C}{ii}] emission extends beyond that of CO. Also, it is observed that generally [\ion{C}{ii}] does not peak where CO peaks but it is offset in the direction of the ionizing stars. 

\subsection{SEST Observations: CO emission lines}\label{sec:sest}

The observations of the CO emission lines were taken with the SEST (Swedish-ESO Submillimeter Telescope) at La Silla Observatory, Chile, during January and September in 2001. SEST was a single dish antenna of 15 meter of diameter operating in the range of frequencies 70$-$365~GHz. It was decommissioned in 2003 with the beginning of operations at the APEX (Atacama Pathfinder Experiment) telescope.

We mapped the central part of the N11 star-forming region in three different regions, which are listed in Table~\ref{tab:sest}. We observed simultaneously the CO~J$=$1--0 and CO~J$=$2--1 lines at the frequencies 115~GHz and 230~GHz with a full width beam at half maximum (FWHM) of 45\arcsec(11~pc linear size) and 23\arcsec(6~pc linear size), respectively. The spatial sampling of the telescope pointings for both frequencies is 24\arcsec. We note that our CO(2--1) observations are not fully-sampled. However, this does not have a significant impact for the goal of this paper (see \S~ \ref{ss:molclouds}). The system temperature varied between 360 and 713~K at 115~GHz and between 235 and 467~K at 230~GHz, during the observations.

The observations were done in position switch mode with a fixed reference off position free of CO emission. A narrow band AOS high-resolution (HRS) spectrometer with 2000 channels, 80 MHz bandwidth and 41.7 kHz channel separation (corresponding to 0.055 km~s$^{-1}$ for the $^{12}$CO~J$=2-1$ line) was used as back end. Intensity calibration was done using the standard chopper-wheel technique. The pointing accuracy, checked during the observations on RDor, was better than 2\arcsec.

   \begin{figure*}\label{fig:co21em}
   \centering
   \includegraphics[width=13.5cm]{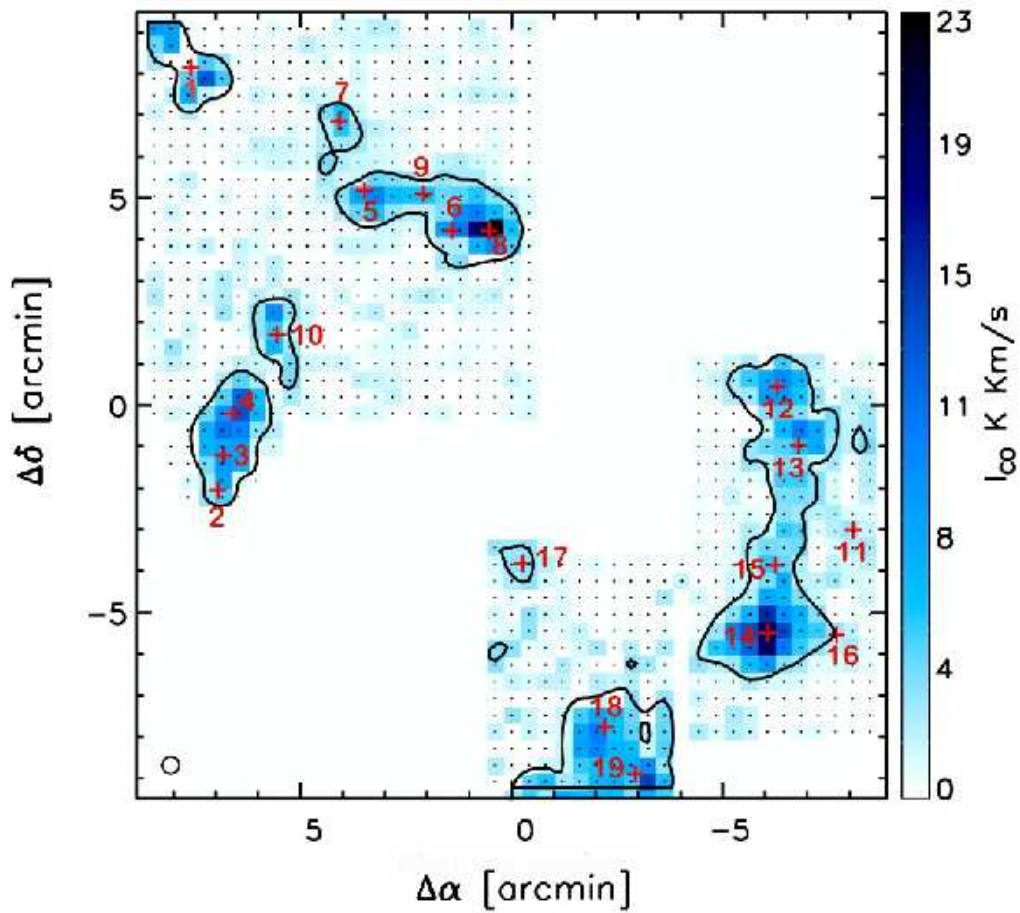}
         \caption{Integrated CO(2--1) emission line of the N11 star forming region, observed with the SEST. Contour corresponds to 2.5 K~km~s$^{-1}$. The dots represent the observed positions. Red crosses mark the position of the molecular clouds listed in Table~\ref{tab:nub}. Offset positions are relative to $\alpha:$$04^{h}56^{m}42^{s}.47$, $\delta:$$-66^{\circ}28^{\prime}42\farcs0$ J2000.0.} \label{fig:co21}
   \end{figure*}

\begin{table}[h]
\begin{minipage}[t]{\columnwidth}
\caption{Characteristics of the SEST observations in the N11 star-forming region.}
\label{tab:sest}
\centering
\renewcommand{\footnoterule}{} 
\begin{tabular}{ccccc}        
\hline\hline                 
\multirow{2}{*}{Zone} & \multicolumn{2}{c}{Coordinates} & $\sigma_{\rm CO(2-1)}^a$  & $\sigma_{\rm CO(1-0)}^b$ \\  
& R.A. & DEC.& \multicolumn{2}{c}{K}   \\  
\hline                       
  Map 1 & $04^{h}57^{m}25^{s}.00$ & $-66^{\circ}24^{\prime}00\farcs0$ & 0.25 & 0.30 \\
  Map 2 & $04^{h}55^{m}36^{s}.00$ & $-66^{\circ}30^{\prime}00\farcs0$  & 0.22 & 0.33 \\
  Map 3 & $04^{h}56^{m}25^{s}.00$ & $-66^{\circ}36^{\prime}00\farcs0$ & 0.20 & 0.27 \\
\hline                                   
\end{tabular}
\end{minipage}
$^a$ computed within the 0.055~${\rm km~s^{-1}}$ channel width. \\
$^b$ computed within the 0.11~${\rm km~s^{-1}}$ channel width.
\end{table}

The data reduction was done with CLASS (Continuum and Line Analysis Single-dish Software\footnote{http://www.iram.fr/IRAMFR/GILDAS}). The complete set of spectra were visually inspected one-by-one, we did not find any anomalies in the spectra. We fitted and subtracted a baseline of second order to each spectrum to correct for residual atmospheric and instrumental emission. The observed  velocity resolution is 0.11 and 0.055~km~s$^{-1}$, for CO(1--0) and CO(2--1), respectively. Finally, to obtain the radiation temperature of the sources, we corrected for the antenna efficiency $\eta$$=$0.7 and 0.5 for CO(1--0) and CO(2--1), respectively \citep{johansson98}. The flux measurements have a photometric calibration uncertainty of 20\%. Table~\ref{tab:sest} lists the 1$\sigma$ noise in a 0.11 and 0.055~km~s$^{-1}$ channel band, for CO(1--0) and CO(2--1) respectively, in each line emission and mapped zone. These values were estimated as the average value of the standard deviation computed in 20 line-free channels at velocities of about 259~km~s$^{-1}$.

\begin{figure*} \centering
\includegraphics[width=16cm,trim = 22mm 0mm 50mm 0mm]{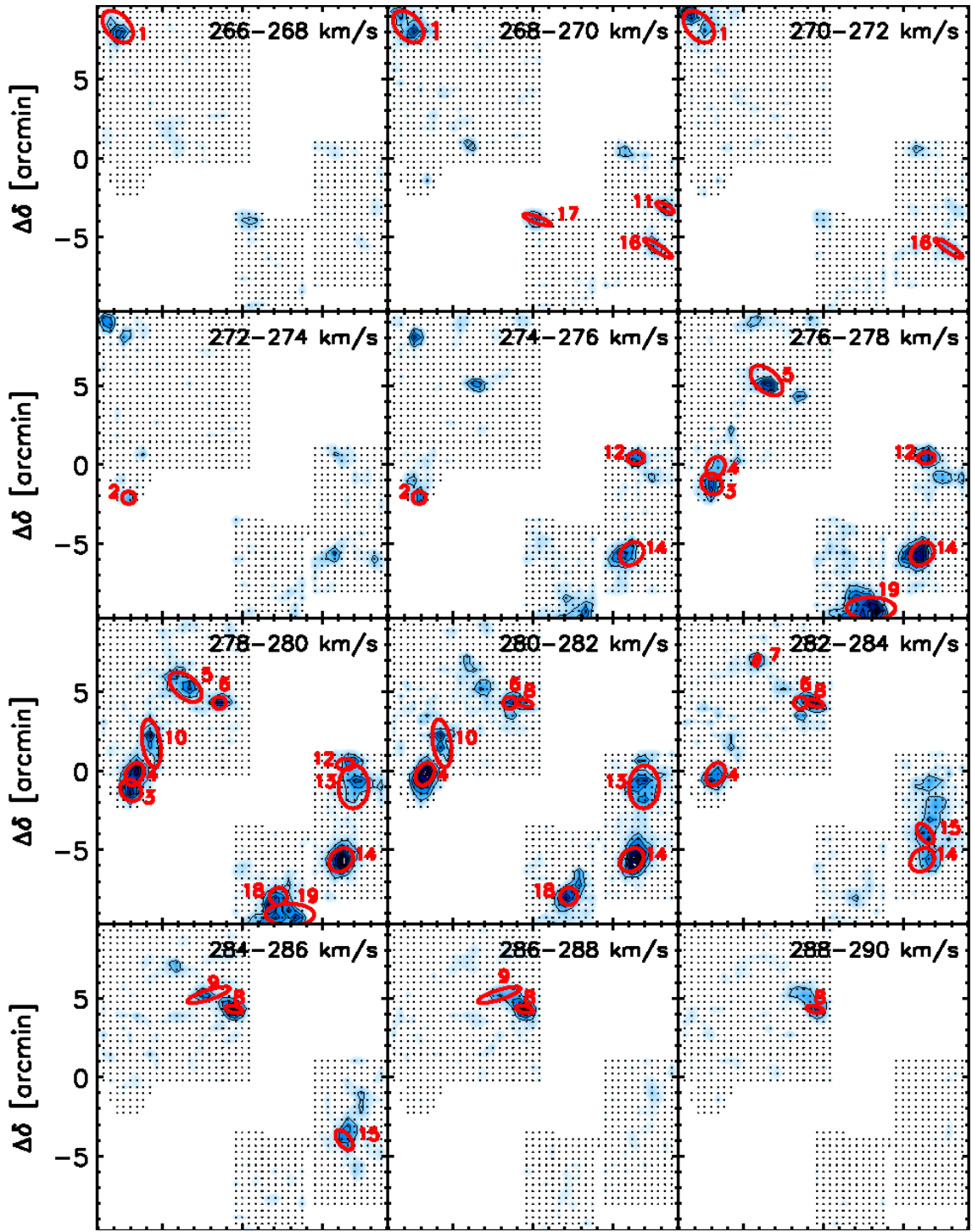}
\vspace{1cm}
\caption{Channel maps for the N11 region. Contours illustrate the CO emission in the transition J$=$2--1 smoothed over 4 pixels. The first contour is 0.75 K km s$^{-1}$, with a spacing of 0.8 K km s$^{-1}$. From the top-left to the bottom-right image, we present channel maps starting from 266 km s$^{-1}$ to 290 km s$^{-1}$, with a velocity width of 2 km s$^{-1}$. Offset positions are the same as in Fig.~\ref{fig:co21em}. We mark in red the molecular clouds identified with CPROPS.} \label{fig:chan}
\end{figure*}

Figure~\ref{fig:co21} shows the integrated CO(2--1) line emission $I_{\rm CO}$, in units of K~km~s$^{-1}$. This image was obtained by integrating the CO emission in a velocity window defined in the total spectrum, from 265 to 289~km~s$^{-1}$. Figure~\ref{fig:chan} presents the CO(2--1) channel maps integrated in bins of 2~km~s$^{-1}$. From these channel maps we can distinguish several components at different velocities.

For each CO transition line, we computed the total emission by adding all the spectra in the field and integrating the emission in the velocity range  265$-$289~km~s$^{-1}$. The total emission corresponds to 1416 K~km~s$^{-1}$ and 1630 K~km~s$^{-1}$ for CO(1--0) and CO(2--1), respectively.
   
\subsection{SIMBA observations: millimeter continuum emission at 1.2~mm}\label{sec:12mm}

To measure the dust emission in N11, we observed the continuum emission at 1.2~mm. This observation was obtained with the SEST IMaging Bolometer Array (SIMBA) at La Silla Observatory in October 2002. SIMBA was a 37 channel hexagonal array operating at 250~GHz. The bandwidth in each channel was about 90~GHz. The Half Power Beamwidth (HPBW) of each element was 24\arcsec.

\begin{figure}[!h]
   \centering
\includegraphics[width=11cm]{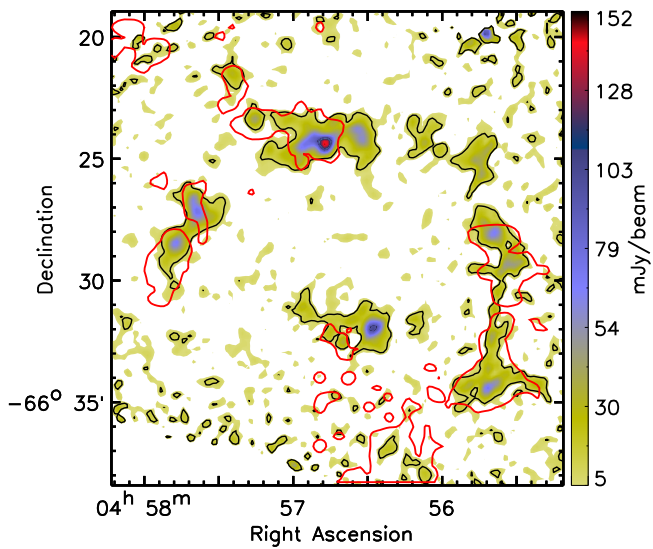}
      \caption{Millimeter continuum emission at 1.2~mm of the N11 star forming region, observed with the SIMBA. Black contour represents the emission at 3$\sigma$$=$12~mJy~beam$^{-1}$. Red contour is the CO(2--1) emission at 2.5~K~km~s$^{-1}$.}\label{fig:12mm}
   \end{figure}

The observations covered the central and northern parts of the N11 complex. The individual maps were produced using the fast scanning mode. The elimination of the correlated sky noise, the co-addition of individual maps, and the photometry was done using the MOPSI package\footnote{MOPSI is a data reduction software package for IR and radio data developed by R. Zylka, IRAM, Grenoble, France}. The $\tau$ zenith opacity was determined performing skydips every 2 hours during the day, and every 3 hours during the night. The flux determination was done for each individual observing run by observing Uranus. The flux measurements in the final coadded map have a photometric calibration uncertainty of 20\%. The sensitivity of the final image is 4~mJy~beam$^{-1}$ (1$\sigma$). Figure~\ref{fig:12mm} shows the observed emission of N11 at 1.2~mm, with CO(2--1) contours overlaid.  

This image was used to measure the millimeter continuum flux of the molecular clouds in N11, identified with the CO observations. We measure the emission in an elliptical aperture defined by the molecular cloud extent at the position of the central coordinates of the cloud. We subtracted the nearby sky emission measured next to the source on an empty image area.  The flux errors correspond to the noise level (standard deviation) in units of Jy~beam$^{-1}$, measured locally on an OFF area next to the cloud, multiplied by the area of the beam and by the root square of the number of beams within the aperture size. The final flux errors were obtained using error propagation of the flux uncertainties and the 20\% photometric calibration errors.

\subsection{Ancillary data}

We complement our data set with two published fully reduced images at 8.6 and 4.8 GHz obtained with the ATCA (Australia Telescope Compact Array) in February 2003 \citep{dickel05}. These observations are part of two surveys of the LMC. The angular resolution is 22\arcsec and 35\arcsec (FWHM beam width) at 8.6 and 4.8 GHz, respectively. The flux uncertainties for the observations are 5\%. 
To measure the fluxes of these radio emission images, we did aperture photometry in the same way as for the emission at 1.2~mm.

%

\section{Molecular clouds in N11}  

In this section, we present the observational results obtained from the SEST observations in N11. 

\subsection{Identification of molecular clouds}\label{ss:molclouds}

The identification of the molecular clouds in N11 was done using the CO(2--1) line emission observations, which has an angular resolution twice that of the CO(1--0) line emission. 
We use the CPROPS algorithm \citep{rosolowsky06} to identify the individual molecular clouds. This program has been used in several other studies of molecular clouds, including the LMC \citep[e.g. ][]{fukui08,hughes10,wong11}. CPROPS is an IDL procedure which identifies clouds from a data-cube using the spectral and spatial structure of the line emission. The physical parameters of the identified molecular clouds are measured as moments of the emission along the spectral and spatial axis. The CPROPS algorithm works as follows. It defines a mask from a $\sigma_{\rm rms}$ value of the data. Within this mask, pixels with emission higher than a threshold value of 5$\sigma_{\rm rms}$, over at least two consecutive velocity channels, are identified as cores of the mask. 
The mask includes pixels with brightness larger than 2$\sigma_{\rm rms}$ emission which meet the two following conditions. They are spatially and spectrally connected to the core pixels and they are not spectrally isolated (i.e. at least two consecutive channels are above 2$\sigma_{\rm rms}$). Every core is decomposed into candidates molecular clouds. The parameters used for this decomposition were: minimum area of 4 beams, line-widths larger than 1.5~km~s$^{-1}$, and a minimum ratio between the flux of the cloud at the peak and at its edge of 3$\sigma_{\rm rms}$. The algorithm defines elliptical molecular clouds, which are extrapolated to the 0~K isosurface to estimate what we would measure with no noise. For a detailed explanation of CPROPS, we direct the reader to \citet{rosolowsky06}.

\begin{figure}[ht]  
\centering
\includegraphics[width=9cm]{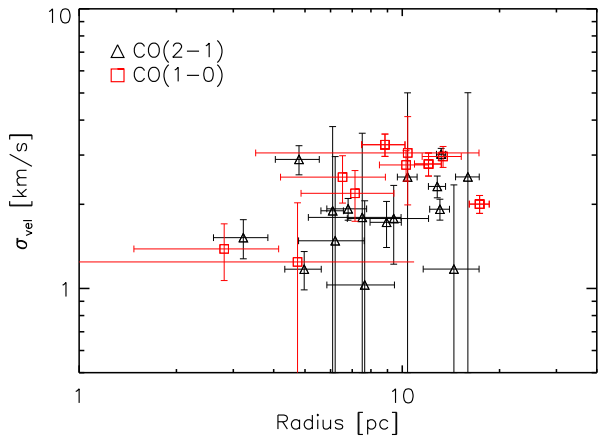}
\caption{Comparison between the radius and velocity dispersion, for molecular clouds obtained from CPROPS runs the CO(1--0) (red
squares) and CO(2--1) (black triangles) data set. The error bars correspond to the 1-sigma errors computed by CPROPS.}\label{fig:under}
\end{figure}

To increase the SNR in our data, we smoothed the data in the velocity axis to a spectral resolution of 0.5~\kms. CPROPS identified 22 individual molecular clouds. We discarded three clouds with low signal-to-noise ratio ($<$4) and which areas are close to our 4 beams threshold. Our CO(2--1) data is not fully-sampled. It was observed at about twice the Nyquist frequency (\S~\ref{sec:sest}). 
 However, the undersampling of the observations do not impact the conclusions of this paper. We quantify the impact of the undersampling on the parameters of the clouds identified by CPROPS with a simple test. We run CPROPS for our fully sampled CO(1--0) (angular spacing of 24\arcsec\ and beam size of 45\arcsec) data cube with a minimum cloud area of one CO(1--0) beam. The results of this run can be directly compared with the CPROPS run on the CO(2--1), for which we have used a minimum cloud area of four CO(2--1) beams. 
 In Figure~\ref{fig:under}, the cloud radii and velocity dispersions for both runs on the CO(1--0) and CO(2--1) data-sets are compared. The error bars are 1-sigma errors on the parameters computed by CPROPS. Although the two cloud parameters do not match exactly one-to-one, the two sets of points are statistically close to each other. There is no systematic difference in radii, and only a slight difference by a factor of 1.2  between the mean velocity dispersion for the CO(1--0) and CO(2--1) lines. We conclude that the undersampling of the CO(2--1) observations has a minor impact on the paper conclusions which are all of statistical nature.  The difference in line widths implies a factor 1.5 systematic on our estimates of the virial masses. We do not correct our 2--1 line widths for this factor but consider it as a systematic uncertainty in our scientific analysis.

Figure~\ref{fig:co21} presents the positions of the molecular clouds in the N11 region. Channel maps in Figure~\ref{fig:chan} show the boundaries of the clouds. Table~\ref{tab:nub} lists the final set of individual molecular clouds identified with CPROPS and their properties. The radii correspond to the geometrical mean of the ellipse semi-axis. CPROPS gives velocity dispersions which we convert to line-widths at half maximum with the relation $\Delta v = \sigma_{\rm v}\,2\,\sqrt{2{\rm\, ln(2)}}$. Uncertainties on the radii, line-widths and CO luminosities come from the fitting procedure in CPROPS \citep{rosolowsky06}. Error bars on the CO luminosity also include the 20\% uncertainty on the calibration. Clouds \#1 and \#5 present different velocity components and might be separated into 2 clouds each. The mean local standard of rest velocity $V_\mathrm{LSR}$ is 278~\kms. 

The total intensity of the identified molecular clouds corresponds to 61\% and 50\% of the total emission (see \S~\ref{sec:sest}) for the CO(1$-$0) and CO(2--1) emission lines, respectively.

\begin{table*}
\begin{center}
\begin{tabular}{ccccccccccc}
\hline \hline
\noalign{\smallskip}
\multirow{2}{*}{Cloud} &  $\delta$R.A. & $\delta$DEC & Velocity & Radius   & $\Delta$v & $L_\mathrm{CO(2-1)}$ & \multirow{2}{*}{$R_{2-1/1-0}$} &  $M_{\rm CO}^{\dagger}$& $M_{\rm vir}^{\dagger\dagger}$ &  \multirow{2}{*}{$M_{\rm vir}/M_{\rm CO}$} \\
\cline{2-3}
\cline{9-10}
\noalign{\smallskip}
 &  \multicolumn{2}{c}{arcmin} & km s$^{-1}$  & pc  &km s$^{-1}$ & 10$^2$ K km s$^{-1}$  pc$^2$ & & \multicolumn{2}{c}{10$^3~M_{\odot}$}  &\\
\hline
\noalign{\smallskip}
 1	 &        7.8	 &        8.3	 &  270	 &   13.2$\pm$0.5	 &    7.1$\pm$0.3	 &    42$\pm$8	 & 0.8$\pm$0.2	 &       52.0$\pm$10.5 &      125.2$\pm$11.2	 &   2.4$\pm$0.5 \\
 2$^{a}$ &   7.1	 &       -2.1	 &  275	 &    $<$5.6		 &    3.9$\pm$0.9	 &     4$\pm$ 1	 & 2.7$\pm$0.5	 &        1.5$\pm$0.4	 &       $<$17	 &  $<$11.2\\
 3	 &        7.0	 &       -1.2	 &  278	 &    9.4$\pm$3.7	 &    4.2$\pm$1.3	 &    21$\pm$11	 & 1.6$\pm$0.3	 &       13.1$\pm$6.6	 &       31.0$\pm$18.2	 &   2.4$\pm$1.8 \\
 4	 &        6.8	 &       -0.2	 &  280	 &    8.9$\pm$1.0	 &    4.1$\pm$0.7	 &    34$\pm$10 & 1.0$\pm$0.2	 &       33.5$\pm$10.3 &       27.6$\pm$10.9	 &   0.8$\pm$0.4 \\
 5	 &        3.6	 &        5.3	 &  278	 &   12.8$\pm$1.0	 &    5.5$\pm$0.5	 &    26$\pm$5	 & 1.5$\pm$0.3	 &       18.5$\pm$3.7	 &       71.6$\pm$14.5	 &   3.9$\pm$1.1 \\
 6$^{a}$&   1.5	 &        4.3	 &  279	 &    $<$5.6		 &    4.7$\pm$1.0	 &     6$\pm$2	 & 2.4$\pm$0.5	 &        2.6$\pm$1.0	 &       $<$25	 &   $<$9.5 \\
 7	 &        4.2	 &        7.0	 &  283	 &    3.2$\pm$0.8	 &    3.6$\pm$0.5	 &     7$\pm$1	 & 1.5$\pm$0.3	 &        5.0$\pm$1.0	 &        7.7$\pm$2.5	 &   1.6$\pm$0.6 \\
 8	 &        0.5	 &        4.3	 &  286	 &    4.8$\pm$1.1	 &    6.8$\pm$0.7	 &    27$\pm$6	 & 2.9$\pm$0.6	 &        9.7$\pm$2.1	 &       41.7$\pm$14.0	 &   4.3$\pm$1.7 \\
 9	 &        2.1	 &        5.2	 &  286	 &    9.2$\pm$1.5	 &    5.3$\pm$0.8	 &    12$\pm$3	 & 1.3$\pm$0.3	 &        9.6$\pm$2.4	 &       49.1$\pm$20.0	 &   5.1$\pm$2.5 \\
10	 &        5.7	 &        1.7	 &  280	 &   13.1$\pm$1.1	 &    4.5$\pm$0.3	 &    22$\pm$5	 & 2.4$\pm$0.5	 &        9.1$\pm$1.8	 &       50.2$\pm$8.3	 &   5.5$\pm$1.4 \\
11	 &       -8.3	 &       -3.1	 &  269	 &    5.0$\pm$1.1	 &    2.8$\pm$0.3	 &     5$\pm$1	 & 1.1$\pm$0.2	 &        4.5$\pm$1.0	 &        7.1$\pm$2.5	 &   1.6$\pm$0.7 \\
12	 &       -6.4	 &        0.4	 &  276	 &    6.2$\pm$1.4	 &    3.5$\pm$0.4	 &    10$\pm$3	 & 0.9$\pm$0.2	 &       12.5$\pm$3.3	 &       14.1$\pm$5.2	 &   1.1$\pm$0.5 \\
13	 &       -6.9	 &       -1.0	 &  280	 &   16.0$\pm$1.6	 &    5.9$\pm$0.8	 &    39$\pm$8	 & 1.1$\pm$0.2	 &       45.0$\pm$9.5	 &      104.2$\pm$30.0	 &   2.3$\pm$0.8 \\
14	 &       -6.2	 &       -5.6	 &  279	 &   10.4$\pm$0.9	 &    5.9$\pm$0.5	 &    73$\pm$15 & 0.9$\pm$0.2	 &       78.7$\pm$16.5 &       67.7$\pm$16.9	 &   0.9$\pm$0.3 \\
15	 &       -6.4	 &       -3.9	 &  284	 &    7.5$\pm$3.2	 &    4.2$\pm$1.1	 &     9$\pm$4	 & 0.8$\pm$0.2	 &       12.1$\pm$5.5	 &       25.2$\pm$13.8	 &   2.1$\pm$1.5 \\
16	 &       -7.9	 &       -5.7	 &  271	 &    6.8$\pm$0.7	 &    4.5$\pm$0.4	 &    10$\pm$2	 & 0.7$\pm$0.1	 &       17.2$\pm$3.6	 &       26.2$\pm$5.1	 &   1.5$\pm$0.4 \\
17	 &       -0.3	 &       -3.9	 &  269	 &    6.1$\pm$0.6	 &    4.5$\pm$0.6	 &     9$\pm$2	 & 0.7$\pm$0.1	 &       12.1$\pm$2.6	 &       22.8$\pm$6.6	 &   1.9$\pm$0.7 \\
18	 &       -3.0	 &       -9.1	 &  278	 &   14.4$\pm$2.4	 &    2.8$\pm$0.5	 &    39$\pm$10 & 0.8$\pm$0.2	 &       51.0$\pm$12.6 &       20.7$\pm$8.2	 &   0.4$\pm$0.2 \\
19	 &       -2.3	 &       -7.9	 &  280	 &    7.7$\pm$2.2	 &    2.4$\pm$0.5	 &    12$\pm$4	 & 1.2$\pm$0.2	 &       10.9$\pm$3.3	 &        8.4$\pm$4.3	 &   0.8$\pm$0.5 \\
\hline
\end{tabular}
\caption{Molecular clouds in N11 identified in CO(2$-$1) with CPROPS. Offset positions are relative to $\alpha:$$04^{h}56^{m}42^{s}.47$, $\delta:$$-66^{\circ}28^{\prime}42\farcs0$ J2000.0. The radii correspond to the geometrical radii for each elliptical cloud. Uncertainties include the error-bars in the CPROPS fitting and 20\% error in the luminosity due to calibration. $^{a}$Clouds unresolved by CPROPS, we use the beam size as upper limit. $^{\dagger}$Molecular masses estimated from the CO(2$-$1) luminosity, scaled to CO(1$-$0) by the $R_{2-1/1-0}$ values, using the $X_{\rm CO}$-factor for the LMC estimated by \citet{hughes10} of $4.7\times 10^{20} {\rm cm^{-2}~(K~km~s^{-1})^{-1}}$ (see section~\ref{sec:mco}). $^{\dagger \dagger}$Virial masses computed with the formula $M_{\rm vir}=190\,\Delta v^2\, R$ for a spherical cloud with a density profile $\propto r^{-1}$ \citep{maclaren88}.}\label{tab:nub}
\end{center}
\end{table*}

\subsection{Comparison with previous data sets}

The N11 star-forming region was previously observed in CO(1--0) as part of the ESO-SEST Key Programme \citep{israel03} with an angular resolution of 45\arcsec. They identified 14 molecular clouds in the same area that we cover. We compare our list with their results. All of the clouds previously identified are in our list of clouds. Our CPROPS run could not resolve two of their clouds (\#16 and \#17, Table~\ref{tab:i03}), due to the undersampling of the observations. These clouds are small ($\sim$30\arcsec, i.e. 7~pc, of radius) and close to each other (40\arcsec, i.e. 10~pc, less than twice the sampling of our observations). Thanks to the higher J$=$2--1 CO resolution, we find six additional clouds. Table~\ref{tab:i03} lists the correspondence between both studies, comparing radii, line-widths and virial masses. Even though there are differences in the data-sets and methodologies, global parameters of the molecular clouds are similar.  Statistically, our line-widths are larger than those of \citet{israel03}, which yields larger virial masses. We checked with the data that this difference mainly comes from the way the line-widths are measured. Those reported by \citet{israel03} are measured on the spectrum at the emission peak of the molecular clouds, while our values are computed on the spectrum integrated over the entire molecular clouds. The higher sensitivity and resolution of our data could also contribute to this difference. We may have included weaker velocity components located in the same line-of-sight of the clouds, which were not detected before.

We also compare our results with those of the MAGMA survey in the whole LMC (CO(1--0) observations, FWHM$=$45\arcsec) \citep{wong11}. Like us, they identified molecular clouds using CPROPS. The parameters used in their decomposition to identify molecular clouds are a minimum ratio between the flux of the cloud at the peak and at its edge of 2$\sigma$.
 In the same area that we observed, they identified 18 clouds. 15 clouds in their and our list coincide. Two of the remaining three clouds are part of a single cloud in our list, while the third one is small ($<$5~pc) and was not detected by our CPROPS run.  Statistically, our CO(2--1) molecular clouds present line-width slightly broader, by a median factor of 1.3.  If we take into account the 1.2 factor which comes from the undersampling of the CO(2--1) observations -- offset value between the mean velocity dispersions of the CO(1--0) and CO(2--1) clouds presented in Section~\ref{ss:molclouds} --, this factor increases slightly to 1.5. 
 A comparison of our molecular clouds with those identified across the entire LMC \citep{wong11} is displayed in Figures~\ref{fig:corr1} and \ref{fig:corr2}, which show that the N11 star-forming region is a singular region in the LMC.

\begin{table}
\begin{center}
\begin{tabular}{p{0.36cm}ccccccc}
\hline \hline
\noalign{\smallskip}
\multicolumn{4}{c}{This work} & \multicolumn{4}{c}{\citet{israel03}} \\
\hline
\noalign{\smallskip}
Ref.  & $R$ & $\Delta v$ & $M_{\rm vir}$ & Ref. & $R$ &$\Delta v$ & $M_\mathrm{vir}$ \\
  \#    &    pc   & {\small km~s$^{-1}$}  &    {\small 10$^3$\msun}   & \#  &  pc & {\small km~s$^{-1}$}  &{\small 10$^3$\msun} \\
\hline
\noalign{\smallskip}
	1	& 13.2	&	7.1	&	125.2	&	16		& 7.7		&	5.8	&	49		\\
		& 		&		&			&	17		& 8.6		&	1.7	&	4.5		\\
	2	& $<$5.6	&	3.9	&	$<$17 	&	$-$      	& $-$	&	$-$	&	$-$		\\
	3	& 9.4		&	4.2	&	31.0  	&	$-$		& $-$	&	$-$	&	$-$		\\
	4	& 8.9		&	4.1	&	27.6  	&	15		& 10.7	&	3.8	&	29		\\
	5	& 12.8	&	5.5	&	71.6  	&	11 		& 7.7		&	4.5	&	30		\\
	6	& $<$5.6	&	4.7	&	$<$25	&	$-$ 		& $-$	&	$-$	&	$-$		\\
	7	& 3.2		&	3.6	&	7.7		&	13 		& $<$5	&	3.1	&	$<$9		\\
	8	& 4.8		&	6.8	&	41.7  	&	10 		& 7.4		&	6.1	&	49		\\
	9	& 9.2		&	5.3	&	49.1  	&	$-$ 		& $-$	&	$-$	&	$-$		\\
	10	& 13.1	&	4.5	&	50.2 		&	14 		& $<$5	&	4.0	&	$<$15	\\
	11	& 5.0		&	2.8	&	7.1 		&	$-$		& $-$	&	$-$	&	$-$		\\
	12	& 6.2		&	3.5	&	14.1  	&	 2		& 7.3		&	2.7	&	10		\\
	13	& 16.0	&	5.9	&	104.2  	&	1		& 10.6	&	2.7	&	15		\\
	14	& 10.4	&	5.9	&	67.7 		&	4		& 11.2	&	5.7	&	69		\\		
	15 	& 7.5		&	4.2	&	25.2  	&	3		& 11.3	&	2.9	&	18		\\
	16	& 6.8		&	4.5	&	26.2 		&	$-$		& $-$	&	$-$	&	$-$		\\
	17	& 6.1		&	4.5	&	22.8  	&	 9		& 7.0		&	2.6	&	9		\\
	18	& 14.4	&	2.8	&	20.7  	&	6		& 19.9	&	2.5	&	24		\\
	19	& 7.7		&	2.4	&	8.4		&	 7		& $<$5	&	1.9	&	$<$4		\\
\hline
\end{tabular}
\caption{Comparison between molecular clouds detected in CO(2$-$1) and previous clouds detected in CO(1$-$0) by \citet{israel03}. To compare the masses, we scale the $M_{\rm vir}$ values from \citet{israel03} to use the formula for spherical clouds with a density profile $\propto r^{-1}$. }\label{tab:i03}
\end{center}
\end{table}

\subsection{Line ratio $R_{2-1/1-0}=I_\mathrm{CO(2-1)}/I_\mathrm{CO(1-0)}$}

\begin{figure}[ht]  \centering
   \includegraphics[width=9cm]{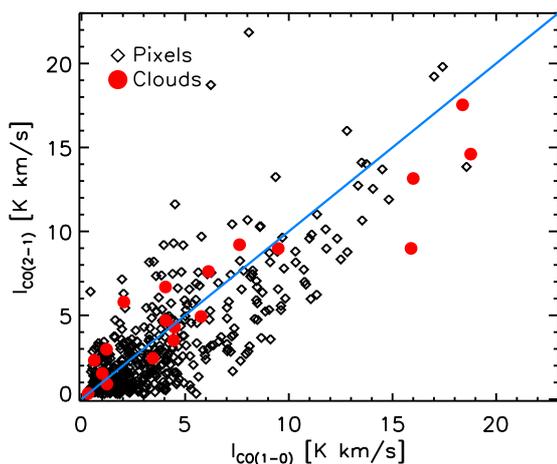}
      \caption{$I_{\rm CO(1-0)}$ vs. $I_{\rm CO(2-1)}$ line emission for each pixel in N11 (black diamonds) and for each molecular cloud in Table~\ref{tab:nub} (red circles). The cloud fluxes were divided by 7. The line corresponds to the median value of the ratio $R_{2-1/1-0}$$=$1.0.}\label{fig:rat}
      \end{figure}

As described in \S 2, we mapped the central part of N11 simultaneously in CO(1--0) and CO(2--1) line emission. From these lines, we compute the ratio of line intensities in K~km~s$^{-1}$, $R_{2-1/1-0}$. To compare these lines, we need to have the data at the same spatial resolution. Thus, we convolve the CO(2--1) data set (FWHM$=$23\arcsec) with a gaussian kernel to the 45\arcsec\ resolution of the CO(1--0) data. Maps of integrated emission for CO(1--0) and CO(2--1) are obtained similarly as described in \S~\ref{sec:sest}, by integrating the emission within a spectral window in the velocity range 265-289~\kms. From these two maps, we derive a $R_{2-1/1-0}$ ratio map for the entire region. We estimate the average $R_{2-1/1-0}$ ratio for each molecular cloud by measuring the emission within an aperture defined by the size of the molecular clouds as identified by CPROPS. Uncertainties on this ratio include the 20\% photometric uncertainties. Table~\ref{tab:nub} lists these values. 
Figure~\ref{fig:rat} shows the CO(1--0) and CO(2--1)  intensity relation for each pixel in our observations of the N11 region and for each molecular cloud. The median value for the $R_{2-1/1-0}$ ratio is 1.0. This is close to the median value of 1.2 reported by \citet{israel03b} for another sample of molecular clouds in the LMC. In Figure~\ref{fig:rat} we observe some dispersion of the pixels around the line. There are a few outliers.
Four molecular clouds, 2, 6, 8 and 10, show large $R_{2-1/1-0}$ values. Clouds 6 and 8 are associated with the embedded IR stellar cluster in N11 identified by \citet{barba03}.

\subsection{Correlations between physical parameters}
     
\begin{figure}[ht]
\centering
      \includegraphics[width=20cm]{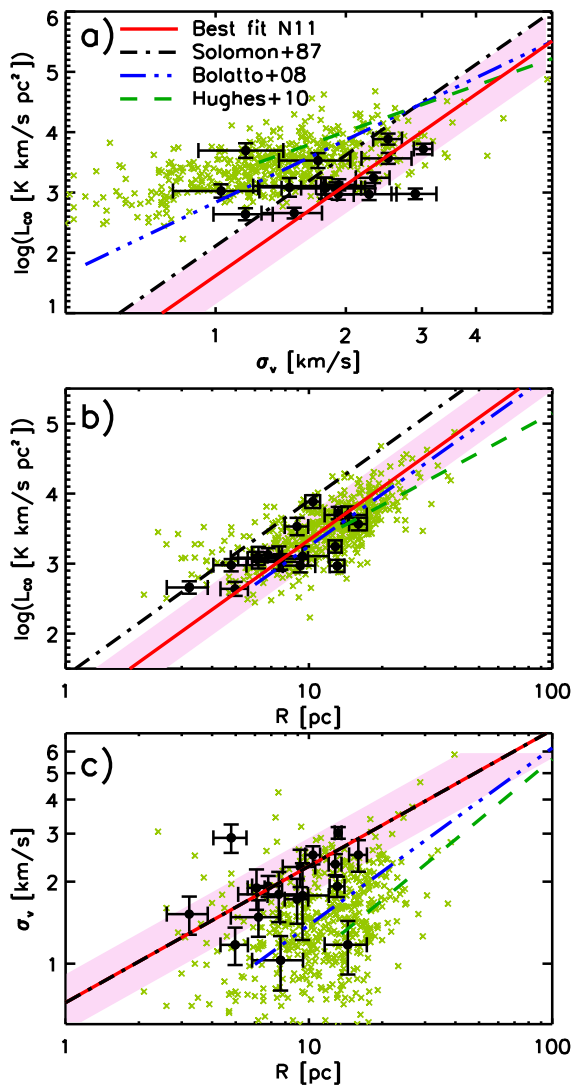}  
      \caption{Correlation graphics. a) log($L_{\rm CO}$) vs. $\sigma_{\rm v}$, b) log($L_{\rm CO}$) vs. $R$ and c) $\sigma_{\rm v}$ vs. $R$.  Straight red lines: best fits to our data, using the Galactic exponents. The shadow area represent the uncertainties for individual data points. Black dot-dashed lines: Galactic relations measured by \citet{solomon87}. Green dashed lines: correlations for GMCs in the LMC from the MAGMA survey by \citet{hughes10}. Blue dot-dashed lines: correlation for dwarf galaxies from \citet{bolatto08}. Green crosses represent the molecular clouds identified in the MAGMA survey of the LMC by \citet{wong11}.}\label{fig:corr1}
\end{figure}

In this section we compare the physical properties of the N11 clouds with empirical relations derived from wider surveys in the Galaxy and the LMC.
We determine the correlations between the physical properties of the molecular clouds $L_{\rm CO(1-0)}$, $R$, $M_\mathrm{vir}$  and $\sigma_{\rm v}$ presented in Table~\ref{tab:nub}. $L_{\rm CO(1-0)}$ values are estimated as the $L_{\rm CO(2-1)}$ measures scaled by the $R_{2-1/1-0}$ estimates. Since the range of the molecular clouds luminosities in N11 is limited, we cannot estimate reliable values for the exponents of the power law correlations between the different parameters. Thus, we fit the N11 relations with the exponents found by \citet{solomon87} for GMCs in the Galactic molecular ring. The multiplicative factor of the following equations result from an error-weighted fit of all data points and the error bars are the dispersion of the individual points,
\begin{eqnarray}  \label{eq:rel1}
L_{\rm CO}&=&6.9_{-4.0}^{+9.0}~R^{2.5}, \\  \label{eq:rel2}
L_{\rm CO}&=&42_{-28}^{+81}~\sigma_{\rm v}^{5},\\ \label{eq:rel3}
M_{\rm vir}&=&91_{-42}^{+79} ~L_{\rm CO}^{0.81}, \\ 
\sigma_{\rm v}&=&0.72_{-0.15}^{+0.19}~R^{0.5}. \label{eq:rel4}
\end{eqnarray}

\begin{figure*}[ht]
	\centering
	\includegraphics[width=12.5cm]{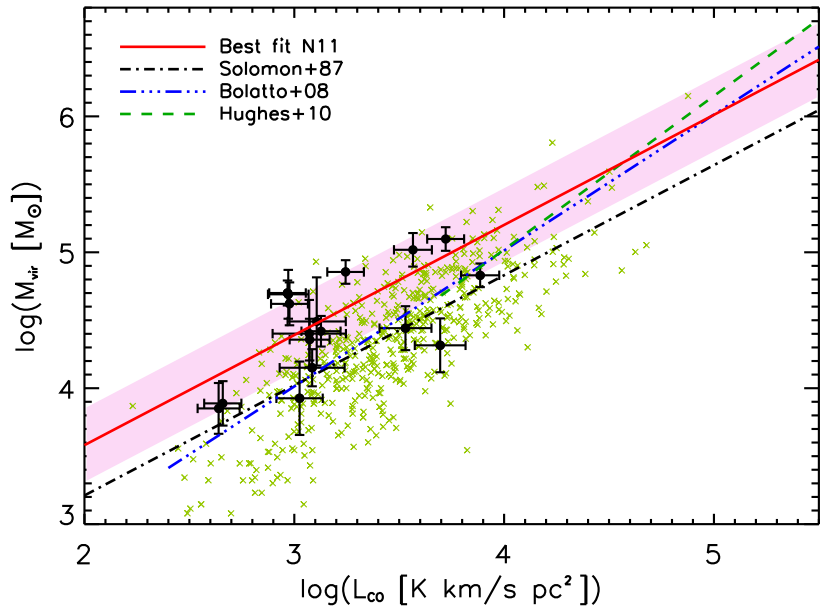}
	\caption{Correlation between $M_{\rm vir}$ and $L_{\rm CO}$ for the molecular clouds of the N11 region. Lines are the same as in Fig.~\ref{fig:corr1}.}\label{fig:corr2}
\end{figure*}

Figure~\ref{fig:corr1} shows the relations $L_{\rm CO}$ vs. $\sigma_{v}$, $L_{\rm CO}$ vs. $R$ and $\sigma_{v}$ vs. $R$. Figure~\ref{fig:corr2} presents the relation between the virial mass and the CO luminosity. In these figures, we compare the N11 molecular clouds with LMC clouds identified by \citet{wong11} as part of the MAGMA survey (green crosses). The best fits for our clouds, using the Galactic exponents, correspond to the solid lines and the shadow areas to the 1$\sigma$-error in the multiplicative factor. We also include, in segmented lines, the relations obtained by \citet{solomon87} for the Galactic molecular clouds, \citet{bolatto08} for GMCs from dwarf galaxies and \citet{hughes10} for GMCs in the LMC. 

\section{Millimeter dust emission at 1.2~mm}\label{sec:mm}

 In this section, we estimate the emission of the dust in N11 at 1.2~mm combining the SIMBA (FWHM$=$24\arcsec), SEST (FWHM$=$23\arcsec) and ATCA 8.6~GHz (FWHM$=$22\arcsec) observations. To do that, we convolved the SEST and ATCA images to the spatial resolution of the SIMBA observations.

In a molecular cloud, the continuum emission at 1.2~mm, $S^{\rm total}_{\rm1.2mm}$, is the sum of free-free and dust radiation. Our observations obtained with the SIMBA bolometer (see \S~\ref{sec:12mm}) provide us with the continuum emission at 1.2~mm. The CO(2--1) emission line falls in the SIMBA bandwidth. Thus, to measure the dust emission at 1.2~mm for the N11 molecular clouds,  the free-free and CO line contribution have to be subtracted.  

We use centimeter-wave observations (at 8.6~GHz) to estimate the contribution from free-free to the 1.2~mm emission. Continuum emission at centimeter wavelengths is composed by two contributions, thermal emission from \ion{H}{ii} gas (free-free emission), and synchrotron emission. Since the spectral index of the thermal emission is flatter than that of the synchrotron emission and the synchrotron emission is not correlated with GMCs, we assume that the emission at 8.6~GHz is mainly tracing the free free emission. We scale the radio continuum emission at 8.6~GHz to 250~GHz (1.2~mm). For an optically thin medium, as the emission at sub-mm/mm wavelengths, the free-free emission depends on the frequency as $S_{\nu}\propto \nu^{-0.1}$. From this relation, we estimate the free-free emission at 1.2~mm as:
\begin{equation}\label{eq:ff}
S^\mathrm{ff}_\mathrm{1.2mm} = \left( \frac{250~\mathrm{GHz}}{8.64~\mathrm{GHz}}\right) ^{-0.1}~S_\mathrm{8.64\,GHz}.
\end{equation}
Uncertainties on $S^\mathrm{ff}_{\rm 1.2mm}$ are computed as the quadratic sum of the errors, including the flux calibration uncertainty (5\% of the flux in the convolved and scaled image), and the uncertainty on the free-free contribution. The latter was estimated comparing, for each cloud, the emission at 4.8 and 8.6~GHz smoothed to a common resolution of 35\arcsec. The error is the absolute value of the difference between the 1.2~mm free-free flux estimated from the 8.6~GHz observation and that computed from the geometric mean value of the 4.8 and 8.6~GHz fluxes.

We estimate the contribution of the CO(2--1) emission line to the SIMBA bolometer as follows. This contribution can be written as $S_\mathrm{CO(2-1)}$$=$$2\mathrm{k}\nu^{3}\mathrm{c}^{-3} \Omega I_\mathrm{CO}/\Delta v$ where $\nu$$=$230~GHz is the frequency of the CO(2--1) line, $I_{\rm CO}$ the emission line intensity, $\Omega$ the solid angle of the Gaussian beam of SIMBA in steradian, and $\Delta v$$=$90~GHz the frequency width of the bolometer. We obtain an expression that we use to compute the contribution of the CO(2--1) line and its error bar:

\begin{equation}\label{eq:co}
\left[ \frac{S^\mathrm{CO}_\mathrm{1.2mm}}{\mathrm{mJy~beam^{-1}}}\right] \simeq 0.20~\left[ \frac{I_\mathrm{CO}}{\mathrm{K~km~s^{-1}}}\right].
\end{equation}

\begin{figure*}
   \centering
   \includegraphics[width=6.2cm,angle=90,viewport=290 0 520 680,clip]{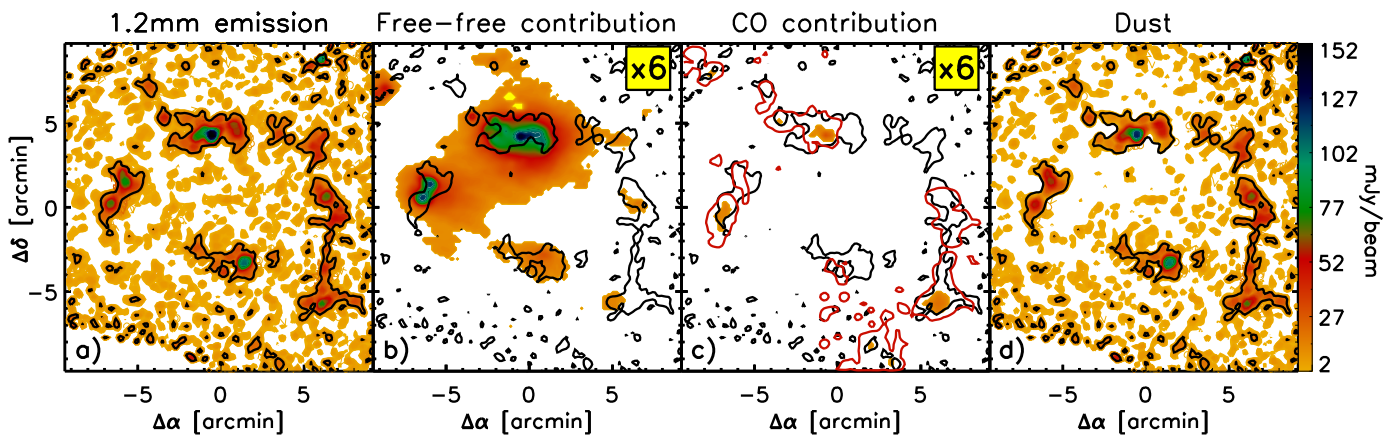}
      \caption{Different contributions to the emission at 1.2~mm. a) Total emission observed with SIMBA. b) Free-free emission at 1.2~mm estimated from the ATCA image at 8.6~GHz, as showed in Eq.~\ref{eq:ff}. c) CO(2--1) contribution to the 1.2~mm emission as observed through the SIMBA bolometer array (see Eq.~\ref{eq:co}). The red contour corresponds to the CO(2--1) emission at 0.5~mJy~beam$^{-1}$. d) N11 dust emission at 1.2~mm obtained after subtraction of the free-free and CO contributions to the total 1.2~mm emission.
All the images are in units of mJy beam$^{-1}$. Intensities of the images b) and c) were amplified by a factor 6. In all the images, the black contour correspond to 3$\sigma$ (12~mJy~beam$^{-1}$) of the total 1.2~mm emission. }
         \label{fig:12mmcont}
   \end{figure*}

To obtain the dust emission, we subtract the free-free and CO contributions from the total continuum emission at 1.2~mm. The error on the dust emission is the quadratic sum of three errors: the error on the 1.2~mm flux including the 20\% uncertainty on the photometric calibration, the error on the free-free subtraction and the error on the CO line contribution. Figure~\ref{fig:12mmcont} shows the different contributions to the emission at 1.2~mm in the N11 region. From the left to the right images: the total emission at 1.2~mm, the free-free emission at 1.2~mm estimated from the ATCA emission at 8.6~GHz, the CO(2--1) line contribution to the total 1.2~mm emission observed with SIMBA, and the resulting dust emission. 
Table~\ref{tab:mm} lists the fluxes contributions to the total 1.2~mm emission for each molecular cloud, measured in the images (convolved to the same spatial resolution) by aperture photometry using the molecular cloud sizes computed by CPROPS and subtracting the nearby background within an aperture of similar size.

%

\section{Molecular masses}\label{sec:mass}

In this section, we discuss and compare three methods to estimate the masses of the molecular clouds based on the CO luminosity, the virial theorem and the millimeter dust emission.

\subsection{Masses from CO luminosity, $M_{\rm CO}$}\label{sec:mco}
In the Galaxy, the mass of a molecular cloud is traditionally estimated from its CO emission. 
The ratio between the CO(1$-$0) integrated intensity $W_{\rm CO}$ and the \hh\ column density \nhh, is the $X_{\rm CO}$-factor ($N({\rm H}_{2})=X_{\rm CO}~W_{\rm CO}$). This conversion factor is uncertain because it depends on the cloud properties. In the Milky Way, it has been calibrated with $\gamma$$-$ray emission \citep[i.e.][]{bloemen86}, the assumption of virial equilibrium \citep[i.e.][]{solomon87}, and dust extinction \citep[i.e.][]{lombardi06}. These empirical studies yield values in the range 2$-$4$\tten{20}$ cm$^{-2}$~(K~\kms)$^{-1}$.   Most recent values from $\gamma-$ray in the Gould Belt are lower \citep{abdo10}, but in average in the Milky Way the studies agree with the previous range of values \citep{bolatto13}.

Several studies \citep[e.g., ][]{cohen88,israel97,fukui08,hughes10} have shown that the Galactic conversion factor does not apply to molecular clouds in the LMC, where the estimated values are between 2 and 4 times higher than the Galactic one. 
In this work, we use the $X_{\rm CO}$-factor computed from the molecular clouds observed and identified by the MAGMA survey in the LMC \citep{hughes10} which corresponds to $4.7\times 10^{20} {\rm cm^{-2}~(K~km~s^{-1})^{-1}}$. To estimate the masses based on the CO luminosity ($M_{\rm CO}$) by using the $X_{\rm CO(1-0)}$ factor estimated from the MAGMA survey, we scale the $L_{\rm CO(2-1)}$ luminosities by the $R_{2-1/1-0}$ values, both listed in Table~\ref{tab:nub}. The inferred CO luminosity masses are also listed in Table~\ref{tab:nub} and they range from 2 to 80~$\times 10^3~M_{\odot}$.

\subsection{Masses from the virial theorem, $M_\mathrm{vir}$}\label{sec:mvir}

The virial masses are computed from the virial equilibrium between gravitational and turbulent kinetic energies. This implies that the molecular clouds are gravitationally bound, ignoring external pressure and magnetic energy. For a spherical cloud with  a profile density of $\rho \propto r^{-1}$, the virial mass is $M_\mathrm{vir} =9/2~R \sigma_{\rm v}^{2}/G$, where $R$ is the radius, $\sigma_{v}$ the velocity dispersion of the gas in 1 dimension, and $G$ the gravitational constant \citep{maclaren88}:

\begin{equation}
\frac{M_\mathrm{vir}}{[M_\mathrm{\sun}]} = 190 \left( \frac{R}{[\mathrm{pc}]}\right ) \left(\frac{\Delta v}{[\mathrm{km\,s^{-1}}]}\right)^{2}.
\end{equation}

To compute $M_{\rm vir}$ we use the line-width and the size estimated from the CO(2--1) emission, both listed in Table~\ref{tab:nub}. The resulting $M_{\rm vir}$ values are listed in the same table. These masses vary between 7 and 130 $\tten{3}~M_{\sun}$. The 1$\sigma$ uncertainty corresponds to the quadratic sum of the errors on the line-width and the radius. In Table~\ref{tab:nub}, we also list the ratio between $M_{\rm vir}$ and $M_{\rm CO}$. The median value for this ratio is 1.9 with a median absolute deviation of 0.9. This ratio increases to 2.8 if we include the offset in the velocity dispersion due to the undersampling (\S~\ref{ss:molclouds}). The median value of the ratio between the virial mass and $L_{\rm CO(1-0)}$ is 21 $M_{\odot}~{\rm (K~km~s^{-1}~pc^2)^{-1}}$ (31 $M_{\odot}~{\rm (K~km~s^{-1}~pc^2)^{-1}}$ including the effect of the undersampling on the velocity dispersion). The ratios for all clouds, but clouds \#14, 18 and 19, are within 3$\sigma$ of this median ratio.

\subsection{Masses from the millimeter dust emission, $M_\mathrm{mm}$}\label{sec:mmm}
The masses of the molecular clouds can also be estimated from the millimeter emission of the dust. In this method, we assume that the dust properties in N11 are the same as those in the Galaxy. There is still an open debate about this hypothesis, since the far-IR dust properties in the Galaxy and in the LMC may differ \citep{galliano11,planck11a17}.

We approximate the spectral dependence of the dust emission with a grey body. Assuming that the emission is optically thin, the dust optical depth $\tau_{d}(\nu)$ can be written in terms of the dust column density. Then, the gas mass estimated from the dust millimeter emission is:
\begin{equation}\label{eq:mmm}
M_\mathrm{mm} = \frac{S_{\nu}\,D^{2}}{B_{\nu}(T_\mathrm{d})\kappa_\mathrm{d}(\nu)\,x_{\rm d}},
\end{equation}
with $S_{\nu}$ the millimeter cold dust emission, $D$ the distance to the LMC, $B_{\nu}$ the black body emission at a dust temperature $T_\mathrm{d}$, $\kappa_\mathrm{d}$ the absorption coefficient per unit dust mass and $x_\mathrm{d}$ the dust-to-gas mass ratio. We can estimate $\kappa_\mathrm{d}$ from the emissivity per hydrogen $\epsilon_{\rm H}(\nu)$ as $\kappa_{\rm d}(\nu)$$=$$\epsilon_{\rm H}(\nu) (x_\mathrm{d}\mu m_\mathrm{H})^{-1}$, where $\mu$ is the mass per hydrogen atom of 1.36. Our masses derived from dust emission, like the masses derived from CO luminosity (in section~\ref{sec:mco}) and the virial masses (in section~\ref{sec:mvir}), include the contribution from helium $\mu=1.36$.  We estimate $\epsilon_{\rm H}({\rm 1.2mm})$ using the recent results obtained with the Planck satellite. By combining Planck and IRAS observations in the Galaxy, \citet{planck11a24} fitted the spectral energy distribution of the local ISM with a grey body function, finding $T_{\rm dust}$$=$17.9 K and $\beta$$=$1.8. They derived a dust emissivity at 250~$\mu$m which agrees with that obtained from COBE observations \citep{boulanger96}. From these Planck results, we compute the dust emissivity per hydrogen atom at 1.2~mm as $\epsilon_{\rm H}$(1.2mm)$=$6.1$\tten{-27}$~cm$^{2}$. Furthermore, \citet{planck11a25} showed that the emissivity $\tau_{250 \mu m}/N_{\rm H}$ increases by a factor of two from the atomic to the molecular gas in the Taurus molecular cloud. Thus, for the dust-to-gas mass ratio in the solar neighborhood of $x_{\rm d}=0.007$  \citep{draine07}, the absorption coefficient for dust in molecular clouds is $\kappa_{\rm d}$(1.2mm)$=$0.77~cm$^{2}$~g$^{-1}$. This value is similar to that used by \citet{bot07} to compute the millimeter masses of the molecular clouds in the SMC. 
 
To compute the millimeter masses, we use a single value of 20~K  for the dust temperatures. This mean value is derived from the temperature map of \citet[][their Figure~7]{planck11a17} at the position of the N11 region. The same figure shows that, around N11, the dust temperature ranges between 15 and 25~K. We will use this range as the maximum uncertainty on the temperatures.

We assume that the dust-to-gas mass ratio scales with metallicity \citep{issa90,james02}. Thus, $x_\mathrm{d}$(LMC)$=$$x_{\rm d}(\odot)~Z_{\rm LMC}$$=$3.5$\tten{-3}$ for $Z_{\rm LMC}$$=$$0.5~Z_{\odot}$. We obtain an expression for the gas mass from the millimeter dust emission at 1.2~mm and for a dust temperature $T_{\rm d}$$=$20~K:
\begin{equation}
\left[\frac{M_\mathrm{mm}}{M_{\odot}} \right] = 158 \left[\frac{S_{\rm 1.2mm}^{\rm dust}}{\mathrm{mJy}} \right],
\end{equation}
where $S_{\rm 1.2mm}^{\rm dust}$ is the cold dust millimeter continuum emission in mJy at 1.2~mm. The multiplicative factor varies from 236 for $T_{\rm d}$$=$15~K to 119 for $T_{\rm d}$$=$25~K.  This variation in temperature gives a maximum uncertainty on the millimeter masses of a factor of 1.5.

The resulting millimeter masses are listed in Table~\ref{tab:mm}. All clouds but one, \#12, are within 2$\sigma$ of the median value $M_{\rm vir}/M_{\rm mm}=1.9\pm0.4$, where the error is the median absolute deviation. Cloud \#12 is an outlier; it is compact, barely resolved by the observations and corresponds to a peak in the millimeter emission.  We discard five molecular clouds because their estimated dust emission, after subtraction of the free-free emission, has a SNR lower than 2$\sigma$. These clouds are in the north-east of N11, where the free-free emission is high (see Figure~\ref{fig:12mmcont}). Their $M_{\rm vir}/M_{\rm mm}$ ratio does not agree with the median value.

\begin{table*}
\begin{center}
\begin{tabular}{cccccccc}
\hline \hline
\noalign{\smallskip}
\multirow{2}{*}{Cloud} &  $S_{\rm 1.2mm}$ & $S_{\rm ff}$ & $S_{\rm CO}$ & $S_{\rm dust}$ & $M_{\rm mm}$ & $M_{\rm vir}$ &  \multirow{2}{*}{$M_{\rm vir}/M_{\rm mm}$} \\
\cline{2-7}
\noalign{\smallskip}
 &  \multicolumn{4}{c}{mJy} &  \multicolumn{2}{c}{10$^3~M_{\odot}$}  & \\
\hline
\noalign{\smallskip}
 1	&  74$\pm$21	 & 13$\pm$6	 & 23$\pm$5	 &   27$\pm$23	 &     $-$		& 125$\pm$11	&  $-$ \\
 3	&  46$\pm$13	 & 8$\pm$4	 & 12$\pm$6	 &   23$\pm$15	 &    $-$		&  31$\pm$18	&  $-$\\
 4	& 166$\pm$47 	 & 21$\pm$10	 & 19$\pm$6	 &  123$\pm$49 &   20$\pm$8	&  28$\pm$11	&   1.4$\pm$0.8 \\
 5	& 106$\pm$30	 & 36$\pm$17	 & 14$\pm$3	 &   53$\pm$35	 &    $-$		&  72$\pm$15	&  $-$\\
 7	&  24$\pm$7	 & 0.2$\pm$0.1	 &  4$\pm$1	 &   19$\pm$7	 &     3$\pm$1	&   8$\pm$3	&   2.5$\pm$1.3 \\
 8	& 212$\pm$60	 & 36$\pm$17	 & 15$\pm$3	 &  159$\pm$62 &  25$\pm$10	&  42$\pm$14	&   1.7$\pm$0.9 \\
 9	& 144$\pm$41	 & 64$\pm$30	 &  7$\pm$2	 &   68$\pm$51	 &     $-$		&  49$\pm$20	&  $-$\\
10	& 401$\pm$114 &125$\pm$59	 & 12$\pm$3	 & 261$\pm$123&  41$\pm$20 	&  50$\pm$8	&   1.2$\pm$0.6 \\
11	&  11$\pm$4	 & 2.3$\pm$ 1.1 &  3$\pm$1	 &    5$\pm$4	 &    $-$		&   7$\pm$3	&  $-$ \\
12	& 142$\pm$40	 &    6$\pm$3	 &  6$\pm$2	 &  129$\pm$41 &   20$\pm$6	&  14$\pm$5	&   0.7$\pm$0.3 \\
13	& 324$\pm$92	 &   10$\pm$5	 & 22$\pm$5	 &  296$\pm$93 &   47$\pm$15	& 104$\pm$30	&   2.2$\pm$0.9 \\
14	& 284$\pm$81	 &   7$\pm$3	 & 40$\pm$8	 &  231$\pm$81 &   37$\pm$13	&  68$\pm$17	&   1.9$\pm$0.7 \\
15	&  93$\pm$27	 & 2.5$\pm$1.2	 &  5$\pm$2	 &   83$\pm$27	 &    13$\pm$4	&  25$\pm$14	&   1.9$\pm$1.3 \\
16	&  54$\pm$16	 & 1.3$\pm$0.6	 &  5$\pm$1	 &   45$\pm$16	 &    7$\pm$3	&  26$\pm$5	&   3.6$\pm$1.6 \\
17	&  24$\pm$8	 & 5.3$\pm$2.5	 &  5$\pm$1	 &   24$\pm$8	 &     4$\pm$1	&  23$\pm$7	&   6.0$\pm$2.9 \\
18	&  84$\pm$26	 &           $-$ 	 & 22$\pm$5	 &   73$\pm$27	 &     12$\pm$4	&  21$\pm$8	&   1.8$\pm$1.1 \\
19	&  38$\pm$12	 & 0.2$\pm$0.1	 &  7$\pm$2	 &   37$\pm$12	 &     6$\pm$2	&   8$\pm$4	&   1.4$\pm$1.0 \\
\hline
\end{tabular}
\caption{Flux contribution to the total 1.2~mm emission for N11 molecular clouds. In the second row we list dust temperatures, obtained by a SED fitting (see section~\ref{sec:mmm}). We separate this emission into 3 different contributions: free-free, CO(2--1) and cold dust emission. Millimeter masses are computed from the cold dust emission (see section~\ref{sec:mmm}). The error bar on dust masses ($M_{\rm mm}$) is derived from the error on $S_{\rm dust}$. It does not include any uncertainty on the dust temperatures. The last column shows the comparison between virial and millimeter masses.}\label{tab:mm}
\end{center}
\end{table*}

\subsection{Comparing the three mass estimates}

The observations provide three independent estimates of the masses of the molecular clouds in the N11 region. Several previous studies have compared the masses derived from CO luminosity with the virial masses for LMC molecular clouds. For the first time, in this paper, we add a third mass estimate based on the millimeter dust emission. Each mass estimate has a large systematic uncertainty, which we can constrain by quantifying what is required for the different methods to yield identical gas masses.

The main uncertainty on the CO masses comes from the $X_{\rm CO}$-factor. The $X_{\rm CO}$-factor is the ratio between the true molecular cloud mass and the CO luminosity,
\begin{equation}\label{eq:xco}
X_{\rm CO} = \frac{M_{\rm true}}{L_{\rm CO (1-0)}}.
\end{equation}

As we pointed out in the introduction, in low metallicity environments the CO molecule is dissociated by the UV radiation from the nearby massive stars. Thus, the LMC $X_{\rm CO}$-factor must be larger than that measured in the Galaxy. However, in the LMC there is a wide range of values reported in the literature, ranging from a factor 2 to 6 higher than in the Galaxy \citep[e.g.][]{cohen88,israel97,fukui08,dobashi08,hughes10}. 

For the virial mass, we assume energy balance between the gravitational binding energy and turbulent kinetic energy. In doing this, we ignore external pressures as well as the magnetic energy. These contributions are difficult to quantify. The importance of these energy terms can vary among the N11 clouds. Some of the molecular clouds could be influenced by the radiation pressure and winds from the near young massive stars. Other clouds could have a high external pressure because they are embedded within a bigger complex. Moreover, the energy terms are computed within the simplifying assumption of spherical geometry. To parametrize the difference between the true mass and the virial mass as estimated in section~\ref{sec:mvir}, we use the $\alpha_{\rm vir}$ parameter  that is the ratio between the kinetic $K$ and gravity $W$ energy terms in the virial equations \citep{mckee92}. For a molecular cloud with a density profile depending on the radius as $\rho\propto r^{-1}$, the $\alpha_{\rm vir}$ parameter is defined as
\begin{equation}\label{eq:alpha}
\alpha_{\rm vir} \equiv \frac{2~K}{|W|} =\frac{9}{2}\frac{R \sigma_{\rm v}^2}{G M_{\rm true}}= \frac{M_{\rm vir}}{M_{\rm true}}.
\end{equation}

The mass estimate based on the dust emission involves uncertainties on the dust temperature and the dust emissivity per hydrogen atom $\epsilon_{\rm H}(\nu)$, which depends on the absorption coefficient of the dust $\kappa_{\rm d}(\nu)$ and the dust-to-gas mass ratio $x_{\rm d}$ (see section~\ref{sec:mmm}). We have computed the masses using a single dust temperature for each cloud but, on scales of a GMC, the dust has a range of temperatures. In Section~\ref{sec:mmm} we estimated the uncertainty on the dust mass associated with the choice of a single temperature to be less than 1.5. The dust emissivity per hydrogen atom depends on two assumptions which are difficult to validate. We assume that the dust-to-gas mass ratio scales as the metallicity and that the millimeter absorption coefficient $\kappa_{\rm d}$(1.2mm) does not change from the Milky Way to the LMC. This is a debated topic \citep{planck11a17,galliano11,draine12} and it is difficult to quantify uncertainties. To parametrize the uncertainties we introduce the factor $\mathcal{K}_{\rm dust} = \kappa_{\rm d}({\rm 1.2mm}) x_{\rm d}/[2.7\times10^{-3}\,{\rm cm}^2{\rm g^{-1}}]$ where the reference value is the value used to estimate the millimeter masses in section 5.3. This is related to the true molecular mass as follows, 
\begin{equation}\label{eq:edust}
M_{\rm true} = \frac{\mathcal{L}_{\rm 1.2mm}({\rm dust})}{B_{\rm 1.2mm}(T_{\rm dust})~\kappa_{\rm d}({\rm 1.2mm}) x_{\rm d}}=\frac{M_{\rm mm}}{\mathcal{K}_{\rm dust}},
\end{equation}
where $\mathcal{L}_{\rm 1.2mm}({\rm dust})$  is the dust luminosity at 1.2~mm per unit frequency, and $M_{\rm mm}$ is the millimeter mass computed in section 5.3.

The measured ratios between $L_{\rm CO}/M_{\rm vir}$ and $M_{\rm mm}/M_{\rm vir}$ constrain the three parameters introduced in formulae~\ref{eq:xco} to \ref{eq:edust} if the three masses are to be identical. In the following equations we use the median values of these two ratios given in section~\ref{sec:mvir} and \ref{sec:mmm}, and the error is the median absolute deviation of these values. Note that these median values do not apply to all clouds. For a small number of clouds, the measured ratios are significantly different from the median values. 

First, the virial mass and the CO mass are equal if
\begin{equation}
X_{\rm CO} = 8.8\pm3.5\, \times10^{20}  {\rm cm^{-2}~(K~km~s^{-1})^{-1}} \, \alpha_{\rm vir}^{-1}, \label{eq:xco_vir}
\end{equation}
and second, the virial mass and the millimeter mass are equal if
\begin{equation}
\frac{\mathcal{K}_{\rm dust}}{\alpha_{\rm vir}} = 0.6\pm0.2, \label{eq:kdu_vir}
\end{equation}
where the error bar does not include the uncertainty on the dust temperature. We assume that the mean temperature of 20~K applies in average to the sample.

 From the previous equation, we estimate the product between the millimeter absorption coefficient and the dust-to-gas mass ratio,
\begin{equation}
\kappa_{\rm d}({\rm 1.2mm})x_{\rm d} =  1.5\pm 0.5~\times 10^{-3} \,{\rm cm^2\,g}^{-1}\,\alpha_{\rm vir}.\label{eq:kx_vir}
\end{equation}

\subsection{Comparison with Gould's Belt clouds in the Milky Way}

\begin{figure}[!ht]
   \centering
  \includegraphics[width=6.3cm,angle=90]{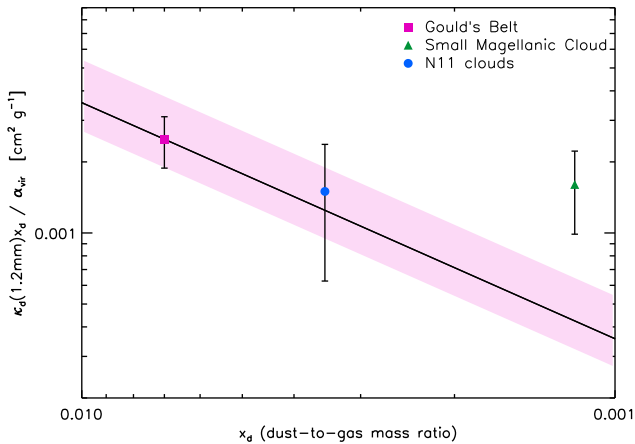}
      \caption{The ratio between the product of the millimeter absorption coefficient $\kappa_{\rm d}$(1.2mm) and the dust-to-gas mass ratio $x_{\rm d}$ with the alpha parameter $\alpha_{\rm vir}$ versus $x_{\rm d}$. This ratio is linearly related to the ratio between the millimeter dust luminosity and the virial mass. The data represent the median values of the $\kappa_{\rm d}({\rm 1.2mm})x_{\rm d}/\alpha_{\rm vir}$ ratio for molecular clouds located in three different environments with different dust-to-gas mass ratio, the Gould's Belt in the Milky Way, several regions in the SMC, and the N11 region in the LMC. The error bar is the standard deviation of each population.  The values for the molecular clouds in the Gould's Belt and the SMC were extracted from \citet{bot07}.  The solid line is the Galactic measurement scaled to different $x_{\rm d}$ values. The shaded area represents the uncertainty on the dust temperature  (from 15 to 25~K).}
         \label{fig:masses}
   \end{figure}
   
In Fig.~\ref{fig:masses} we compare our results for the N11 clouds with earlier measurements done by \citet{bot07} for clouds with comparable masses in the Gould's Belt (i.e. in the solar neighborhood) and the SMC. The figure shows the median ratio $\kappa_{\rm d}({\rm 1.2mm})x_{\rm d}/\alpha_{\rm vir}$ versus the dust-to-gas mass ratio $x_{\rm d}$, for the molecular clouds in the three different environments.  For a given dust temperature, the data points in Fig.~\ref{fig:masses} scale with the ratio  between the dust luminosity at 1.2~mm and the virial masses (Eq.~\ref{eq:edust}). For the Galactic and SMC clouds, these values were computed from the ratio between the millimeter and virial masses listed in Tables~3 and 4 of \citet{bot07}. Error bars are the standard deviation of the individual values. The solid line shows the Galactic value of $\kappa_{\rm d}({\rm 1.2mm})x_{\rm d}/\alpha_{\rm vir}$ scaled to different gas-to-dust mass ratios. The pink region represents the uncertainty on the dust temperature, assuming a typical value of $T_{\rm d}=20$~K which varies between 15 and 25~K. The median value for the N11 clouds agree with the scaled Galactic value. This indicates that the observed difference in the millimeter dust luminosity to virial mass ratio ratio between the Galactic and N11 clouds can be accounted for by the difference in the gas metallicities. This statement dos not apply to the SMC clouds as reported by \citet{rubio04} and \citet{bot07,bot10}. The millimeter dust opacity $\kappa_{\rm d}$ could be higher in the SMC than in the Milky Way. This interpretation is supported by the difference in the sub-mm/mm dust spectral energy distributions \citep{israel10, planck11a17}. \citet{draine12} have proposed that the enhanced millimeter dust emission observed in the SMC could be accounted for by dipolar magnetic emission from ferromagnetic particles.

 \subsection{Discussion}

Our observational results constrain the product between the $X_{\rm CO}$-factor and the alpha virial parameter (Eq.~\ref{eq:xco_vir}), and the ratio between the millimeter dust absorption coefficient and the virial parameter (Eq.~\ref{eq:kx_vir}) for molecular clouds in the N11 region. In this section, we discuss the most likely values for these three parameters.

If the measured virial masses of the N11 molecular clouds are their true cloud masses, our $X_{\rm CO}$-factor is about 3 times larger than the canonical value estimated for Galactic clouds \citep{solomon87}.  {\bf The effect of the undersampling of the CO(2--1) observations will increase by a factor 1.5 the difference between the averaged $X_{\rm CO}$-factor in the Milky Way and our estimate. } Within uncertainties, it agrees with that found with the low-resolution (2\farcm6 half-power beamwidth) NANTEN survey in the LMC by \citet{fukui08}, estimated from the virial and CO luminosity masses. 
But, it is about twice the value estimated by the MAGMA survey in the LMC (see Section~\ref{sec:mco}) and that estimated by most of the studies of extragalactic GMCs, including GMCs from galaxies in the local group \citep{blitz07, bolatto08}, as observed in Figure 7 in \citet{leroy11}, where our $X_{\rm CO}$-factor estimate corresponds to $\alpha_{\rm CO}=19~M_{\odot}~{\rm pc^{-2}~({\rm K~km~s^{-1}})^{-1}}$. The $X_{\rm CO}$-factor for the N11 clouds will agree with these previous estimates in the LMC if the $\alpha_{\rm vir}$ parameter is a factor $\sim$~2 (see Eq.~\ref{eq:xco_vir}). 

For molecular clouds in the LMC, \citet{wong11} compared the $\alpha_{\rm vir}$ parameter\footnote{This value depends on the assumed $X_{\rm CO}$-factor value.} with the CO luminous masses of the clouds (see the right panel of their Fig.~15). Using the canonical value of $X_{\rm CO}$, most of the LMC clouds have $\alpha_{\rm vir}>1$ and half of them $\alpha_{\rm vir}>2$. Using the $X_{\rm CO}$-factor estimated by \citet{hughes10}, half of the clouds have $\alpha_{\rm vir}>1$ and only a few $\alpha_{\rm vir}>2$. 
For molecular clouds in the outer Galaxy, \citet{heyer01} found $\alpha_{\rm vir}$ to be larger than 2 for clouds with CO masses smaller than $3\times 10^3~M_{\odot}$. Most of the N11 clouds are more massive than this limit (see Table~\ref{tab:nub}).  However, in a recent study of galactic GMCs, defined previously by \citet{solomon87} and \citet{heyer09}, \citet{dobbs11} found that Galactic clouds, in the N11 mass range, have $\alpha_{\rm vir}>1$ and about half of them have a $\alpha_{\rm vir}>2$ (see their Fig.~1). This finding is in agreement with numerical simulations of the turbulent ISM in the Milky Way presented in their paper. \citet{dobbs11} claim that there is no need for molecular clouds to be entirely bound to reproduce the observed star formation rate. Star formation will occur in the densest part of the unbound clouds, where gravity wins over turbulence. 

For $\alpha_{\rm vir}=2$, Equation~\ref{eq:kx_vir} yields a dust absorption coefficient for the N11 clouds of $\kappa_{\rm d}=0.86\,{\rm cm^2~g}^{-1}$. This value is similar to the Galactic value presented in Section~\ref{sec:mmm}.  Therefore, we favor these values for the $\alpha_{\rm vir}$ and $\kappa_{\rm d}$ parameters because they do not require the dust emissivity, nor the dynamical state of the clouds, to differ between the LMC and the Milky Way. For $\alpha_{\rm vir}=2$, we also get a $X_{\rm CO}$-factor akin to that estimated by the MAGMA survey in the LMC. The ratio between the virial parameter and the dust far-IR properties is plotted in Figure~\ref{fig:masses}.  The data does not allow us to separate the two parameters $\alpha_{\rm vir}$ and $\kappa_{\rm d}$. For N11 with respect to the Galaxy we favor a variation of $\kappa_{\rm d}$ associated with metallicity and no variation in  $\alpha_{\rm vir}$. However, this interpretation does not apply to SMC clouds. The figure shows that the observed $\kappa_{\rm d}({\rm 1.2mm})x_{\rm d}/\alpha_{\rm vir}$ values for the GMCs in the SMC are above than that estimated for the Galaxy and N11, by a factor of 3 to 4. 

%

\section{Summary}

We have presented high sensitivity and spatial resolution observations of the CO(2--1) line emission observed with SEST, for the N11 star-forming region in the LMC. Simultaneously with this CO transition, we observed the J$=$1--0 transition, which allowed us to measure the ratio between the CO(2--1) and CO(1--0) integrated emission for the entire map. This ratio has a median value of unity for the entire region. We identified 19 molecular clouds using the CPROPS algorithm which yields their CO luminosity, sizes, line-widths and virial masses. We also presented 1.2~mm continuum observations which we use to estimate millimeter dust fluxes. In this paper, and for the first time in the LMC,  we compute molecular masses estimated from the millimeter dust luminosity ($\mathcal{L}_{\rm 1.2mm}({\rm dust})$) and compare them with the masses obtained from the CO luminosity ($L_{\rm CO}$) and virial theorem ($M_{\rm vir}$). The main results of this study are the following ones.

\begin{itemize}
\item The correlations between  CO luminosity, line-widths, sizes, and virial masses in the N11 clouds are in agreement with those found in earlier CO surveys of the LMC, sampling a larger set of clouds and a larger range of cloud masses \citep{fukui08, wong11}. 

\item  The measured ratios $L_{\rm CO}/M_{\rm vir}$ and $\mathcal{L}_{\rm 1.2mm}({\rm dust})/M_{\rm vir}$ constrain the $X_{\rm CO}$-factor and the dust emissivity per hydrogen atom at 1.2~mm as a function of the virial parameter $\alpha_{\rm vir}$ ($M_{\rm vir}/M_{\rm true}$, the ratio between the kinetic and gravity energy terms in the virial equations).

\item The comparison between the CO and virial masses yields a conversion factor of $X_{\rm CO}=8.8\pm3.5~\times10^{20}  {\rm cm^{-2}~(K~km~s^{-1})^{-1}} \, \alpha_{\rm vir}^{-1}$.  The comparison between the virial and millimeter dust luminosity yields a product between the gas-to-dust mass ratio $x_{\rm d}$ and the millimeter absorption coefficient per unit of dust mass $\kappa_{\rm d}$(1.2mm) of $1.5\pm0.5~\times 10^{-3}~{\rm cm^2\,g^{-1}\,}\alpha_{\rm vir}$.

\item The comparison with previous studies of molecular clouds in the LMC suggests that N11 clouds are unbound, as observed in several Galactic molecular clouds within the same range of masses \citep{dobbs11}. The median values for the $\alpha_{\rm vir}$ and $\kappa_{\rm d}$ parameters are 2 and $0.86\,{\rm cm^2~g}^{-1}$, respectively. These values do not require peculiar dust properties nor dynamical clouds properties different from those observed in the Milky Way. 

\item We do not find in N11 a large discrepancy between the dust millimeter and virial masses, as reported for GMCs in the SMC by \citet{rubio04} and \citet{bot07,bot10}. The ratio $\kappa_{\rm d}({\rm 1.2mm})x_{\rm d}/\alpha_{\rm vir}$ in N11 is half of that measured for Gould's Belt molecular clouds, which can be accounted for by a factor two lower gas-to-dust mass ratio. This difference is the same as that observed between the gas metallicities. If the two samples have similar $\alpha_{\rm vir}$ median values, their dust far-IR properties should be also similar. 

\end{itemize}

%

\begin{acknowledgements}
We are grateful to A. Leroy for helping us to compute the properties of the molecular clouds using CPROPS. 
M.R. and C.H. acknowledge support from FONDECYT grant No.1080335. M.R. and F.B. acknowledge the support from the ECOS research grant C08U03. M.R. is supported by the Chilean Center for Astrophysics FONDAP No.15010003. A.D.B. wishes to acknowledge partial support from a CAREER grant NSF-AST0955836,and from a Research Corporation for Science Advancement Cottrell Scholar award.
\end{acknowledgements}

\bibliographystyle{aa}
\bibliography{papers}
\end{document}